# The electron capture cross section using the r-CTMC method: a review and a suggestion for improvements

Fabio Sattin[1]

Consorzio RFX, Associazione Euratom-ENEA
Corso Stati Uniti 4, 35127 Padova, Italy

**Abstract**

In this work we review the classical recipe for estimating partial cross sections for electron capture into selected energy levels within the Classical Trajectory Monte Carlo (CTMC) method [R.L. Becker and A.D. MacKellar, J. Phys. B: At. Mol. Phys. **17** (1984) 3923], and point to a deficiency of the method when applied to the CTMC in its *r*-version, according to the terminology introduced by Cohen [J.S. Cohen, J. Phys. B: At. Mol. Phys. **18** (1985) 1759]. We suggest, in this case, an alternative recipe based on the generation of stationary classical spatial distributions for electrons in arbitrary levels. The output of the CTMC simulations is projected onto these classical distributions rather than on the quantal ones. A comparison between the two approaches is done.

---

[1] E-mail: fabio.sattin@igi.cnr.it

# I. Introduction

The Classical Trajectory Monte Carlo (CTMC) method is one of the most widely used computational tools for modelling heavy particles collisions with electron transfer, due to its simplicity of implementation, accuracy and wide range of applicability. Along the years, several attempts have been made to improve from the original version of the method [1] to get even more accurate results.

Indeed, the CTMC method is made of three steps: 1) the choice of the initial conditions; 2) the solving of the dynamical equations; 3) the determination of the final configuration. Giving an accurate recipe for all of the three steps in problems involving several electrons and/or non-hydrogenlike ions, is a tremendously complicated matter, far from being fully solved. Dealing only with hydrogenlike and fully stripped ions removes complications related to point (2), since the interparticle forces are exactly known, but still leaves some open questions related to points (1) and (3). The problems arise because it is not possible to devise classical statistical distributions that exactly match the corresponding Quantum Mechanical (QM) ones for all of the phase space variables. The original choice, and still the most commonly adopted, is to pick up initial conditions from a microcanonical ensemble. This allows matching exactly QM energy and momentum distribution, at the expense of a poor spatial distribution. This choice is based on the idea that it is the velocity matching between electron and projectile nucleus, more than the spatial overlap between the trajectories, to determine the output of the simulation. There is the further advantage, from the numerical point of view, that microcanonical distributions are, by construction, stationary under the potential of the target nucleus. However, the concern about the deficient description of large-impact-parameter collisions, which is unavoidable within this picture, led several researchers to devise more efficient strategies, able to accurately reproduce even the spatial distribution. Accurate statistical distributions can be useful even for semiclassical variants of the CTMC method [2]. We are referring here to the works by Eichenauer *et al* [3], where the (truncated) Wigner function was used; to Hardie and Olson [4] which adopted a linear combination of microcanonical spatial distributions corresponding to different binding energies. Explicitly, they wrote

$$\rho(\mathbf{r}) = \sum_i w_i \rho_\mu(\mathbf{r}, E_i) \qquad (1)$$

where

$$\rho_\mu(\mathbf{r}, E_i) = \frac{4}{\pi^2 Z^3} E_i^{5/2} \sqrt{\frac{Z}{r} - E_i} \qquad (2)$$

is the spatial microcanonical distribution for an electron bound to a nucleus of charge $Z$, with binding energy $E_i$ (> 0)-Note that, contrary to common usage, we found easier to define throughout this paper the binding energy as a positive quantity: when comparing expressions written here with their counterparts in the literature, the reader must be aware of the possibility of the change of sign: $E \rightarrow -E$. Also, atomic units will be used throughout this paper.

Eq. (2) is normalized so that

$$4\pi \int_0^{Z/E_i} r^2 \rho_\mu dr = 1 \tag{3}$$

The weights $w_i$ and the energies $E_i$ were chosen empirically by Hardie and Olson to get a good representation of the hydrogen $1s$ orbital.

Both approaches were encompassed within Cohen's formulation [5]: Cohen started from the probability distribution function in the phase space $\hat{\rho}(\mathbf{r},\mathbf{p})$, rather than in position space. The phase space density $\hat{\rho}(\mathbf{r},\mathbf{p})$ can be integrated over either of the variables $\mathbf{r}$, $\mathbf{p}$, to get the probability distribution function (PDF) for the other variable. In particular,

$$\int \hat{\rho}(\mathbf{r},\mathbf{p}) d\mathbf{p} = \rho(\mathbf{r}) \tag{4}$$

Cohen postulated that $\hat{\rho}(\mathbf{r},\mathbf{p})$ could be written in terms of just the binding energy: $\hat{\rho}(\mathbf{r},\mathbf{p}) \equiv f(E)$. For a Coulomb potential, the energy $E$ is related to coordinates $\mathbf{r}$, $\mathbf{p}$, through

$$E = -p^2/2 + Z/r \tag{5}$$

hence the l.h.s. of Eq. (4) becomes

$$\int \hat{\rho} d\mathbf{p} = 4\sqrt{2}\pi \int_0^{Z/r} f(E)(Z/r - E)^{1/2} dE \tag{6}$$

Cohen imposed the r.h.s. of Eq. (4) to be equal to its ground state QM value:

$$\rho \equiv \rho_{QM} = \frac{1}{\pi} \exp(-2r) \quad , \tag{7}$$

(where we have explicitly set $Z = 1$, since for the moment we are dealing only with hydrogen atoms) and, by inverting the integral equation (a Volterra equation), found $f(E)$:

$$f(E) = \frac{\exp(-1/E)}{2^{3/2}\pi^{5/2}E^{3/2}} \left[ \frac{1}{4} W_{-1/2,-1/2}\left(\frac{2}{E}\right) - \left(1 - \frac{2}{E}\right) W_{1/2,-1/2}\left(\frac{2}{E}\right) \right] \quad , \tag{8}$$

where $W$ is the Whittaker function. The technical details of the inversion are not relevant for the present paper, and can be found in Cohen's paper. Cohen showed also that the empirically truncated Wigner function used in Eichenauer *et al* [3] is essentially an approximation of his own exact result. We will show later that the same is true for the Hardie and Olson's approach.

Following Cohen's terminology, we will hereafter use the term ``*p*-CTMC'' to refer to the CTMC method in connection with the microcanonical distribution, and ``*r*-CTMC'' to refer to its modified versions. The *r*-CTMC, by construction, allows the spatial distribution to be correctly recovered, and the momentum distribution quite closely reproduced. However, it has also some drawbacks, and in this work we are explicitly addressing the following one: since the distribution of binding energies is not a Dirac delta centered at the QM value, but is spread over a finite support, some of the statistical runs start with electrons that are bound to the target nucleus less (or more) strongly than allowed by QM. This, of course, has consequences over the integrated results (total cross sections), as shown in Cohen's work. But some further subtle effects arise when one looks for differential results, such as partial cross sections for capture into selected quantum levels.

Indeed, the standard procedure for identifying capture into a given quantum level is the binning procedure by Becker and MacKellar [6]. First, one computes the binding energy $U$ of the electron from its coordinates relative to the projectile nucleus (of charge $Z_p$):

$$U = \frac{Z_p}{r} - \frac{p^2}{2} \qquad (9)$$

From Eq. (9), then, a "classical" (real) quantum number $n_c = Z_p/(2U)^{1/2}$ is computed, which is finally "quantized", i.e., put into correspondence with one of the (integer) quantum numbers $n$ allowed by QM. The rule is that $n$ must be chosen such that

$$[n(n-1/2)(n-1)]^{1/3} \leq n_c \leq [n(n+1/2)(n+1)]^{1/3} \qquad (10)$$

We remind that, for more detailed analysis, similar relations have been written down for angular quantum numbers, too, but we will not consider them here. Also, we just mention that this procedure has been recently modified to be more accurate when dealing with non hydrogen-like atoms [7,8].

While well suited for *p*-CTMC, the binning procedure (10) is not consistent with *r*-CTMC. One can easily be convinced of this by applying Eq. (10) to the *initial* electron distribution itself: to obtain backwards consistency, all electrons should be found within to the same level; instead one gets electrons spread over several *n* levels, because *U* varies from run to run. More important, the same thing is expected to happen for the captured electrons: those randomly chosen with a large binding energy to the target will probably populate projectile's lower *n* levels than should be allowed to do (and the converse will be true for those picked from the low-energy part of the distribution). But it should be clear that this effect is not a consequence of the stochasticity intrinsic to the CTMC method, mimicking the QM randomness; instead, is an artefact due to our inconsistent method of binning.

In this work we present a slightly different method of generating the initial distribution. It is a generalization of Cohen's formula to excited states. The real conceptual novelty in the work, however, is that *f(E)* (or, better a generalization of it) is used to identify not only initial distributions, but also final ones. This provides a different method to assign probabilities for capture into selected states. Results are compared against quantum methods and standard CTMC runs using Eq. (10), and commented.

We wish to remark that an analysis of energy probability distribution function in order to correctly recover the number of electron transfers has recently been suggested within the *p-*CTMC approach [9,10].

## II. The algorithm
### II.1. The initial distribution

Using Eq. (2), it is easy to see that Eq. (6) can be written as a weighted average of microcanonical PDFs:

$$\rho(\mathbf{r}) = \int dE\, p(E) \rho_\mu(\mathbf{r}, E) \qquad (11)$$

Of course, the integration runs only over the finite support of $\rho_\mu$ 's (given by eq. 2).
where *p(E)* must fulfil

$$p(E) = \sqrt{2}\pi^3 Z^3 \frac{f(E)}{E^{5/2}} \qquad (12)$$

(see also Eq. 16 in Cohen's paper).

It is interesting to see that this formula can accommodate Hardie and Olson's own, too: by comparing with Eq. (1),

$$p(E_i)\Delta E_i \approx w_i \qquad (13)$$

where $\Delta E_i$ is the energy interval between two energy values $E_i$.

In Eq. (11), *p(E)* is the probability density for picking up a microcanonical basis function with energy *E*. In the l.h.s. of Eq. (11), Cohen inserted the ground state density. In principle however, that equation can be inverted for almost any function $\rho$. We verified this by explicitly performing the inversion for 1*s*, 2*p*, 3*d* hydrogen states. The reason we chose these states is that, wavefunctions of states with angular numbers $l < n - 1$ have nodal points. It is easy to show that, to fit such functions, *p(E)* must be negative somewhere. Indeed, *p(E)* is slightly negative even for the above mentioned states, but the contribution of this "unphysical" part to the total is negligible. Since we assign *p(E)* the meaning of a probability density, it is hard to justify negative values.

Although *p(E)* admits analytical expression for all the states, its explicit form is rather cumbersome, since involves an increasing number of hypergeometric functions. The simplest example is given in Eq. (8) itself. For this reason, we resorted to approximate the *p(E)*'s with more manageable expressions, and found that a suitable choice was the inverse gamma distribution:

$$p(E) = p_\Gamma(E) \equiv \frac{1}{\Gamma(\nu)\beta^\nu E^{\nu+1}} \exp\left(-\frac{1}{\beta E}\right) \qquad (14)$$

where $\beta, \nu$ are two parameters to be determined. As an example, we plot in Fig. (1) both the exact *p(E)*-from Eqns. (8,12) as well as its best fitting from Eq. (14).

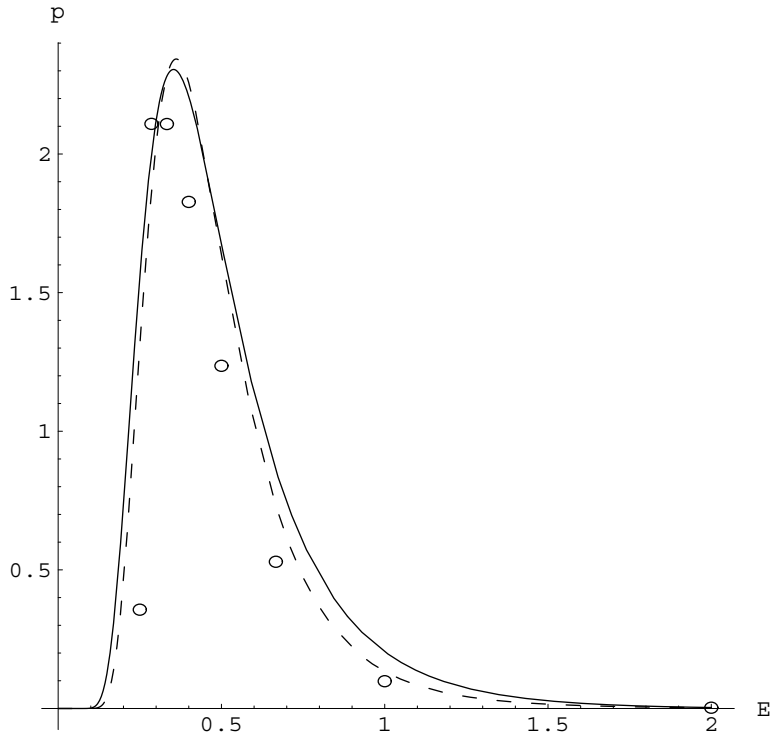

Fig. 1. Dashed line, *p(E)* from Eq. (14); solid line $\sqrt{2}\pi^3 f(E)/E^{5/2}$, with *f(E)* from Eq. (8) ; Circles, empirical fitting coefficients (rescaled according to Eq. 13) from [4]. Coefficients $\beta$, $\nu$ have been chosen so to give the exact mean radius and energy.

Indeed, we found empirically that, as a rule of thumb, good fits were obtained for these choices of β,ν:

$$\beta = \frac{n+1}{5Z^2}$$

$$\nu = 6n$$

Using Eq. (14), the integral in Eq. (11) can be analytically computed for any couple of β,ν:

$$\rho(r) = \frac{4x^{-1/2}}{\pi^2(Z\beta)^5 \Gamma(\nu)} \left[ \Gamma\left(\nu - \frac{5}{2}\right) {}_1F_1\left(-\frac{1}{2}, \frac{7}{2} - \nu, -x\right) + x^{5/2} \sqrt{\frac{\pi}{4}} \frac{\Gamma\left(\frac{5}{2} - \nu\right)}{\Gamma(4-\nu)} {}_1F_1\left(\nu - 3, \nu - \frac{3}{2}, -x\right) \right]$$

(15)

with $x = r/(\beta Z)$, and where $_1F_1$ is the Kummer confluent hypergeometric function. Expression (15) appears intimidating, but several statistical quantities can be analytically computed from it: by example, the first moments

$$< r^k > = 4\pi \int r^k \rho(r) r^2 dr \quad (k = 1, 2) \tag{16}$$

with the result

$$< r > = \frac{5}{8} \beta \nu Z \tag{17}$$

$$< r^2 > = \frac{7}{16} (\beta Z)^2 \nu(\nu + 1) \tag{18}$$

Analogously, one can compute statistical quantities of energy directly using (14), e.g., the mean energy

$$< E >_{\beta\nu} \equiv \int E p_\Gamma(E) dE = \frac{1}{\beta(\nu - 1)} \tag{19}$$

or its second moment

$$< E^2 >_{\beta\nu} = < E >_{\beta\nu}^2 \left( \frac{\nu - 1}{\nu - 2} \right) \tag{20}$$

and, also, the maximum of p(E): $dp_\Gamma(\overline{E})/dE = 0$

$$\overline{E} = \frac{1}{\beta(\nu + 1)} \tag{21}$$

A further insight to the why of the particular choice for *p(E)*(Eq. 14) can be found also in the paper by Kunc [11]: there, he demonstrated that functions very close to $p_\Gamma$ describe the velocity PDF for classical electrons orbiting into central fields.

It is possible to see that the wavefunction (15) differs asymptotically from its quantum mechanical counterpart by a factor $r^{1/2}$. In Fig. (2) we plot the QM ground state density, as well as the result (15).

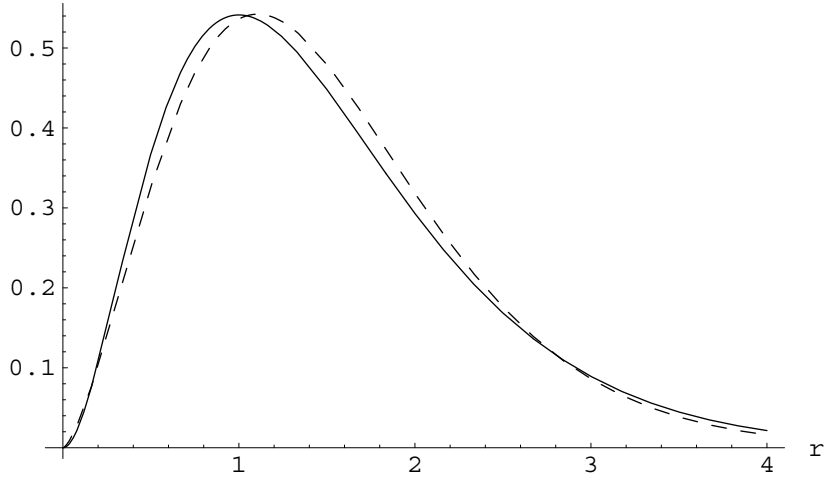

Fig. 2. Radial wavefunction $4\pi r^2 \rho$ for the hydrogen ground state density, $(n, l) = (1,0)$. Solid line, QM result; dashed line, present work. Had we used *p(E)* given by Eqns. (8,12), the exact QM result would have been reproduced (see [5]).

With this we are done for the part concerning the generation of arbitrary initial distributions. The practical recipe to generate a stationary spatial distribution corresponding to a quantum state $(Z, n, l)$ is therefore: first, the parameters $\beta, \nu$, are evaluated from the empirical rule above; second, using these values for $\beta,\nu$ a value $E_0$ of the binding energy is randomly picked up from the distribution *p(E)* (Eq. 14); finally, electron coordinates are chosen with standard methods from the stationary microcanonical distribution $\rho_\mu (r, E_0)$ (Eq. 2). This path is simpler to implement than directly sampling from Eq. (15).

## II.2. The final distribution

Let us suppose of having performed a full CTMC calculation and got as output the PDF $f_{cap}(E)$, yielding the fractions of electrons captured by the projectile with a binding energy *E*. If we were performing a QM calculation, $f_{cap}(E)$ would collapse to a sum of Dirac delta functions:

$$f_{cap}(E) = \sum_i c_i \delta(E - E_i) \qquad (22)$$

with the sum running over quantum levels. Thus, the probability for capture into *i*th level is $P_i = c_i$ .

Since our system obeys Classical Physics, we expect $f_{cap}(E)$ to be instead a smooth continuous function of *E*. Becker and Mackellar's binning procedure allows us, however, to write a classical version of Eq. (22):

$$f_{cap}(E) = \sum_n c_n \Theta_n^{BM}(E) \qquad (23)$$

with

$$\Theta_n^{BM}(E) = \begin{cases} 1 & E_{inf}(n) < E < E_{sup}(n) \\ 0 & \text{otherwise} \end{cases} \qquad (24)$$

The extrema $E_{inf}$, $E_{sup}$ are determined by Eqns. (9, 10).

We have already shown that Eq. (23) is logically inconsistent within the *r*-CTMC formalism. However, we still wish to preserve its appealing features even in this case, *i.e.* we wish to be able to write-at least approximately:

$$f_{cap}(E) \approx \sum_j d_j f_j(E) \qquad (25)$$

The condition that $f_{cap}$ must collapse to Eq. (12) when charge exchange does not occur, leads us to postulate

$$f_{cap}(E) \approx \sum_{\beta,\nu} d_{\beta,\nu}\, p(\beta,\nu,E) \approx \sum_{\beta,\nu} d_{\beta,\nu}\, p_\Gamma(\beta,\nu,E) \qquad (26)$$

In the middle term of Eq. (26) the projection is done over the exact PDF's *p(E)* (Eq. 11), which are then replaced, in the rightmost term, by the approximated $p_\Gamma$ using Eq. (14).

The projection (26) is not as sharp as (22) or (23,24) since the support of functions $p_\Gamma$ overlap for different (β, ν), (β', ν'). However, it is possible to show that the amount of overlapping is negligible if the indices (β, ν), (β', ν') are sufficiently far apart.

The other question is, how do we choose the values (β, ν) in the sum (26) ?

Two answers can be given. The most straightforward one is to exploit the left-to-right arrow in the (*n, l*) ↔ (β, ν) correspondence: since, after all, we are interested in computing captures into quantum levels (*n, l*), we can choose a suitable set of them, and evaluate correspondingly $\beta_{nl} \equiv \beta(n, l)$, $\nu_{nl} \equiv \nu(n, l)$ using the rules in the section above. The coefficients $d_{\beta\nu}$ in Eq. (26) are then found as parameters of the fitting of the empirical data $f_{cap}(E)$ using the $p_\Gamma(\beta_{nl}, \nu_{nl}, E)$ set of functions. This procedure, however, revealed a drawback: we are implicitly assuming that charge exchange populates few (*n,l*) states and *exactly* them, therefore the sum (26) can be reduced to few terms. This is not true: first, classically, captures are spread over a continuum of states; second, even if they were concentrated around few values of energies, there is not classically any constraint to force them to be equal to the QM value. Fixing (β, ν) parameters in (26) and remaining with just the *d's* as fitting parameters, turned into a poor quality of the fit.

We preferred, therefore, the other way round: the parameters β, ν are no longer preassigned, but instead are computed as fitting coefficients, like the *d*'s. With this greater flexibility we are able to get much better fits. Each basis function is thus assigned to an energy level. This point had to be resolved a bit empirically: the connection (β, ν) → (energy level) can be done in several ways. By example, by computing the average energy $<E>_{\beta\nu} = 1/(\beta(\nu-1))$ for each basis function (using Eq. 19), and assigning the whole PDF to the quantum level nearest in energy to $<E>_{\beta\nu}$. We found that more reasonable results were found by using the position of the maximum, $\overline{E} = 1/(\beta(\nu+1))$ (Eq. 21). Indeed, this is not completely arbitrary: the best fit of the exact PDF's (11) by the $p_\Gamma$'s (14) is done when matching the maxima of the curves, not the mean values.

The fraction of captures into that level is, finally, estimated by the corresponding $d_{\beta\nu}$.

### III. Numerical examples

We will test the method over three cases. The system H(1s) + He$^{2+}$ → H$^+$ + He$^+$ *(n,l)* is ideal to test this approach: it is well documented in literature, and the capture to helium occurs into few levels, thus making manageable the projection (26). In fig. (3) we plot the total cross section for this process, computed using a CTMC code *ad hoc* developed according to the above guidelines, thus comparing our results with those already present in literature: a full QM molecular approach [12], a different classical method [13], and a standard *p*-CTMC computation [14]. Our CTMC results underestimate the correct value of σ for impact velocity *v* < 1, but this is a well known deficiency of CTMC methods. More relevant is the perfect agreement between our CTMC results and Olson's one at *v* = √2, which gives confidence about our correct implementation of the code. (The fact that the two results are obtained using r- and p-CTMC models is of no concern here. Cohen already showed that the differences between the two approaches disappear at high *v*).

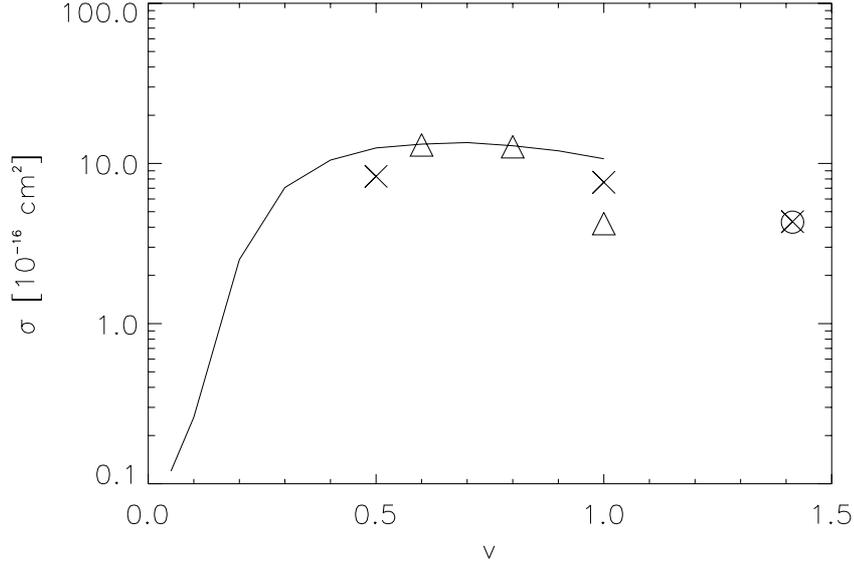

Fig. 3. Total cross section for electron capture from H(1s) by He$^{2+}$ *versus* impact velocity. Solid line, data from [12]; triangles, data from [13]; circle, data from [14]; crosses, present work.

The best way to understand the present approach is by looking at fig. (4): here, we have plotted the number of captures per energy unit at fixed energy. Using standard numerical routines, the CTMC data have been fitted to a linear combination of three curves of the kind (14), whose parameters are given in table I.

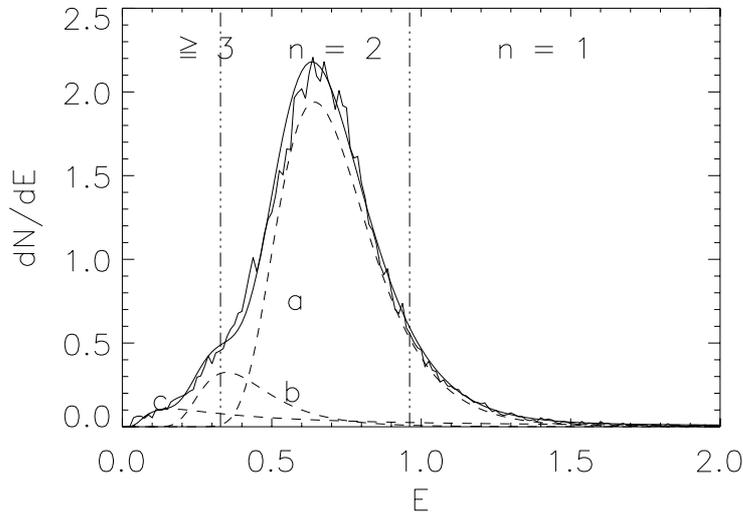

Fig. 4. Differential number of captures *versus* final binding energy, for collision between H(1s) and He$^{2+}$ at impact velocity $v = 1/2$. Broken line, CTMC results; dashed lines labelled with ``a'', ``b'', ``c'', the energy components (Eq. 26) from the fitting; solid smooth line, the overall fitting curve. The vertical chain lines mark the regions where captures pertain to a single quantum level (displayed on top) according to rule (Eq. 10).

| Curve | $d_{\beta\nu}$ | $\beta$ | $\nu$ | $\overline{E}_{\beta\nu}$ | $n_{eq}$ |
|---|---|---|---|---|---|
| a | 0.773 | 0.085 | 17.274 | 0.642 | 1.76 |
| b | 0.123 | 0.301 | 8.519 | 0.349 | 2.39 |
| c | 0.104 | 4.16 | 0.500 | 0.160 | 3.53 |

Table I. Parameters of the fitting curves of fig. (4). $d$ is the weight of each component, $\overline{E}_{\beta\nu}$ is the average energy over that component, $n_{eq}$ is the ``effective'' quantum number: $n_{eq} = (2/\overline{E}_{\beta\nu})^{1/2}$.

About 10% of the curve falls into the region labelled as ``$n = 1$''. With the standard binning procedure (Eq. 10), these captures would pertain to the $He^+(1s)$ state. Instead, with the present method, no captures contribute to this state. Now, we proceed by assigning the fraction of captures pertaining to each component to the quantum energy level whose principal number $n$ is closest to the estimated effective quantum number $n_{eq}$: the results are listed in table II for some values of the impact velocity $v$ and together with estimates from the other models. We have added a point also for $H(1s) + Be^{4+}$ collisions.

| Ion | V | N | PW | SB | Harel98 | Olson81 | Ille99 |
|---|---|---|---|---|---|---|---|
| **$He^+$** | ½ | 1 | 0.0 | 0.1 | 0.0 | -- | -- |
| | | 2 | 0.9 | 0.84 | 0.93 | -- | -- |
| | | ≥3 | 0.1 | 0.06 | 0.07 | -- | -- |
| | 1 | 1 | 0.0 | 0.12 | 0.01 | -- | 0.07 |
| | | 2 | 0.86 | 0.71 | 0.74 | -- | 0.56 |
| | | ≥3 | 0.14 | 0.17 | 0.24 | -- | 0.28 |
| | √2 | 1 | 0.0 | 0.19 | -- | 0.14 | -- |
| | | 2 | 0.29 | 0.49 | -- | 0.52 | -- |
| | | ≥3 | 0.71 | 0.33 | -- | 0.29 | -- |
| **$Be^{3+}$** | 1 | 1 | 0.0 | 0.0 | -- | -- | -- |
| | | 2 | 0.0 | 0.20 | -- | -- | -- |
| | | ≥3 | 1.0 | 0.80 | -- | -- | -- |

Table II: Fraction of captures into levels $n = 1,\ldots, \geq 3$ of $He^+$ and $Be^{3+}$ for different impact velocities. ``PW'' refers to the present method; ``SB'', to the same CTMC calculations, but binned using Eq. (10); ``Harel98'' refers to data from [12], ``Olson81'' from [14]; ``Ille99'' from [13]. The fits were performed using three terms ("basis functions") in the sum (26).

Basically, at the lowest velocities the present algorithm ``drains'' captures from the lowest and highest levels, where the naive binning argument would have placed them. The results are in rough accordance with those found using a QM molecular approach, that should be more reliable in this velocity range. When *v* increases, ionization begins to appear. The curve *dN/dE* does not go to zero any longer as $E \to 0$, but crosses the axis $E = 0$ at a finite value (corresponding to a finite fraction of captures into unbound states with $E < 0$). Fitting this curve becomes increasingly difficult since a larger and larger number of basis function would be needed. An example is shown in fig. (5), where the equivalent of fig. (4) is displayed, but computed at an impact velocity $v = \sqrt{2}$. In order for the fitting routine to converge, the experimental curve had to be artificially truncated below $E = 0.025$. This value is arbitrary, but final results do not appear to be critically dependent from this threshold. It is clear, from Table II, how our algorithm performs worse and worse as *v* increases: at $v = 1$, its performances is already comparable-indeed slightly worse than Becker and MacKellar's-while at $v = \sqrt{2}$ there is no competition.

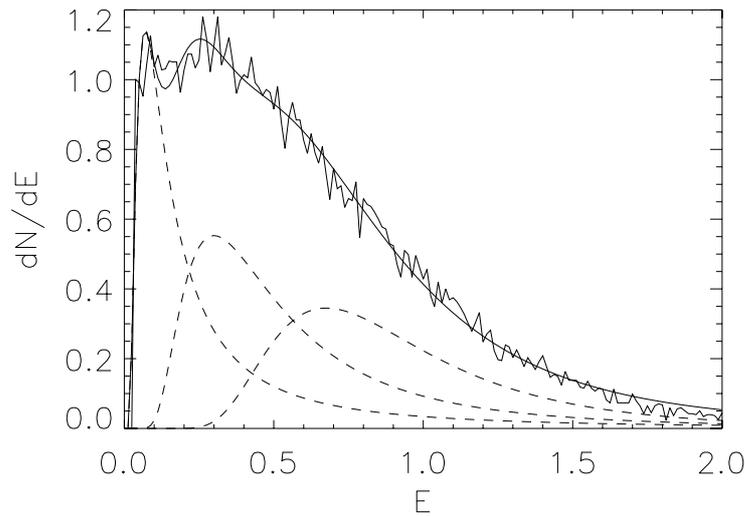

Fig. 5. The same as Fig. 4, but for $v = \sqrt{2}$

On the other hand, as long as ionization can be neglected, even more highly charged ions can be dealt with satisfactorily: in fig. (6) we see that the fitting of the captures H(1s), $Be^{4+} \to Be^{3+}$ is fairly good.

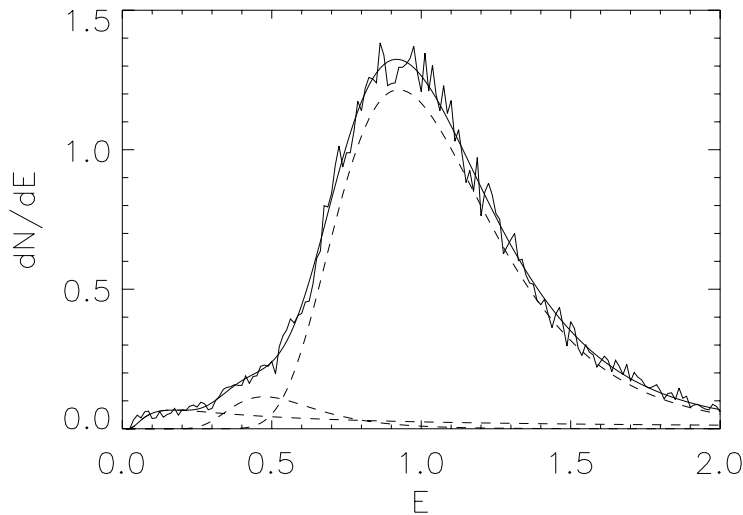

Fig. 6. Differential number of captures into Be$^{3+}$ at collision velocity $v = 1$.

As a third bench test, we consider the scattering He$^+$ + H$^+$ → He$^{2+}$ + H, for which recent partial cross sections have been computed with QM methods [15]

This case revealed fairly tough to handle. Infact, even at low speed (v = ½), the electron-Helium interaction is difficult to be correctly subtracted off the results. However, even these rough estimates, assign about 50% of the captures to the H ground state, to be compared with a 40% from Eq. (10). QM calculations [15], albeit performed at higher energies, assign almost all captures to $n = 1$ state.

## IV. Conclusion

This work consists of two parts: the first is rather trivial in itself, since simply provides an approximate generalized form for the function found by Cohen. It may be important in case one needs to easily generate excited-states' wavefunction within the *r*-CTMC method. However, it acquires importance when one needs to analyze the *output* of a *r*-CTMC simulation. Indeed, we showed that the standard Becker and MacKellar's procedure is not consistent with the logics within the *r*-CTMC method. Therefore, in the second part of this work, we suggested an alternative procedure to compute partial cross section into energy levels which, instead, appears to preserve consistency.

Being logically more satisfactory does not mean being always convenient to practical purposes. Indeed, we must admit that the suggested procedure to get partial cross sections is rather cumbersome: the numerical fitting of $f_{cap}(E)$ is not an easy task. The examples we provided in section III yielded a mixed result (see Table II). There were some improvements: from comparison with QM calculations, it appears that an improvement appears for estimating the fraction of captures into the dominating level. This is, of course, not a surprise, since the method is conceived exactly to squeeze captures into fewer levels.

On the other hand, this procedure is not competitive with the Becker and MacKellar's one (Eq. 10) when the number of basis functions is large. Perhaps, in this case a ``mixed'' approach could be fruitful: a small number of dominant levels (say, 2 or 3) are identified through the method here outlined. These dominant components are subtracted off from the total capture PDF, and the remaining is analyzed using standard technique (Eq. 10). We have applied this approach to the cases described in the previous section, but the corrections to the cross section were too small ($\approx 0.01$) to be meaningful.

For practical purposes, therefore, the main goal of this paper should be its hinting some weak points within the established formalism. The suggested remedies can be looked at as either solutions for important but somewhat niche cases, or suggestion for further possible improvements over similar lines.